\documentclass[fleqn,usenatbib]{mnras}

\usepackage{newtxtext,newtxmath}
\usepackage[T1]{fontenc}
\usepackage{ae,aecompl}

\usepackage{graphicx}
\usepackage{amsmath}
\usepackage{amssymb}
\usepackage[space]{grffile}
\usepackage{multicol}
\usepackage{pdflscape}
\usepackage[normalem]{ulem}

\newcommand*\dd{\mathop{}\!\mathrm{d}}

\title[Correlations in \textit{Gaia} DR2 unbound stars]{Searching for
  correlations in \textit{Gaia} DR2 unbound star trajectories}

\author[Montanari et al.]{
F. Montanari$^{1}$\thanks{E-mail: francesco.montanari@uam.es},
D. Barrado$^{2}$\thanks{E-mail: barrado@cab.inta-csic.es}, and
J. Garc\'ia-Bellido$^{1}$\thanks{E-mail: juan.garciabellido@uam.es}
\\
$^{1}$Instituto de F\'isica Te\'orica IFT-UAM/CSIC, Universidad
  Aut\'onoma de Madrid, Cantoblanco 28049 Madrid, Spain\\
$^{2}$Departamento de Astrof\'{i}sica, Centro de Astrobiolog\'{i}a (CSIC-INTA), ESA-ESAC. Camino Bajo del Castillo s/n.\\$\quad$28692 Villanueva de la Ca\~{n}ada, Madrid, Spain\\
}

\pubyear{2019}

\begin{document}
\label{firstpage}
\pagerange{\pageref{firstpage}--\pageref{lastpage}}
\maketitle

\begin{abstract}
  Scattering events with compact objects are expected in the primordial black hole (PBH) cold dark matter (CDM) scenario due to close encounters between stars and PBH in the dense environments of dwarf spheroidals. We develop a Bayesian framework to search for correlations among Milky Way stellar trajectories and those of globular clusters and dwarf galaxies in the halo, and other nearby galaxies. We apply the method to a selection of hypervelocity stars (HVS) and globular clusters from \textit{Gaia} DR2 catalog, and known nearby (mostly dwarf) galaxies with full phase-space and size measurements. We report positive evidence for trajectory intersection $\sim$20--40~Myr ago of up to 2 stars, depending on priors, with the Sagittarius dwarf Spheroidal (dSph) galaxy when assuming \citet{2018MNRAS.tmp.2466M} distance estimates. We verify that the result is compatible with their evolutionary status, setting a lower bound for the stellar age of $\sim$100~Myr. However, such scattering events are not confirmed when assuming \citet{2019A&A...628A..94A} distance estimates. We discuss shortcomings related to present data quality and future prospects for detection of HVS with the full \textit{Gaia} catalog and Sagittarius dSph.
\end{abstract}

\begin{keywords}
  cosmology: dark matter -- Galaxy: kinematics and dynamics -- Galaxy:
  stellar content -- galaxies: kinematics and dynamics -- galaxies:
  statistics
\end{keywords}



\section{Introduction}

The hierarchical structure formation scenario assumes that large galaxies are formed by mergers of smaller ones, which bring in both gas (hydrogen), stars and dark matter (DM). These smaller structures, generically called dwarf galaxies, orbit around the larger galaxy and interact with it. Some appear tidally disrupted by previous crossings through the disk and are elongated, and others are still approaching it and have more or less spherical shape. All this substructure have large mass-to-light ratios, in some cases larger than 1000, making them extremely difficult to detect in the sky. Their numbers were predicted, within the cold dark matter scenario, to be large, hundreds to thousands of objects orbiting each large galaxy. However, only about a dozen had been observed until SDSS and DES discovered several tens of them~\citep{2015ApJ...813..109D,Newton:2017xqg,Simon:2019ojy}, solving the so-called substructure problem when extrapolated to the whole sky.

The low surface brightness of dwarf galaxies could be explained in the PBH CDM scenario~\citep{Garcia-Bellido:2017fdg} due to the loss of stars via close encounters with massive primordial black holes comprising the dark matter halos of all galaxies. In this scenario, stars in the shallow potential wells of dwarf galaxies are more likely to get slingshot away due to close encounters with DM black holes, acquiring velocities in the hundreds to thousands of km/s~\citep{Clesse:2016vqa}. Such hypervelocity stars (HVS), likely unbound to the Milky Way potential, should travel across the sky and their trajectories should point back to the dSph from which they originate. High-velocity stars are also found in the core of globular clusters (GC) \citep{2012A&A...543A..82L}, which may indicate a population of PBH, and also in this case some of them may acquire a velocity above the escape threshold~\citep{Clesse:2016vqa}.

In this paper we develop a Bayesian framework for the detection of such close encounters via the correlation of stellar trajectories in \textit{Gaia} DR2 catalog~\citep{2016A&A...595A...1G, 2018A&A...616A...1G} with trajectories of dwarf and other nearby galaxies, and GC. If massive black holes are indeed responsible for the depletion of stars from dwarf galaxies, it is expected that a few slingshot events must have happened in the last 100 million years inside dwarf galaxies in the Milky Way. In particular, since the probability of events is proportional to the density of the dwarf galaxy, one expects the most massive ones to be the source of HVS. Unfortunately, \textit{Gaia} has limited resolution for distant stars, and only those relatively close to the Sun are measured with sufficient accuracy in 6D phase space (we consider stars up to $\sim$13 kpc).

Recently, \cite{2018MNRAS.tmp.2466M} found that some of the observed HVS were {\em not} pointing away from the center of the Milky Way, as was naively expected, but rather towards the disk, as if they had originated in the halo of our galaxy. Furthermore, \citet{2018ApJ...866..121H} reported one star whose orbit has non-negligible probability of having passed near the Large Magellanic Cloud in the past. This prompted us to explore the possible origin of HVS and whether they could originate in the dSph that orbit around the MW within a radius of several tens of kiloparsecs, and which could have been travelling for several tens of millions of years from their sources and velocities up to ten times larger than typical stellar velocities in the halo.

We compute the close encounter evidence between HVS and Milky Way GC,
dwarf, and nearby galaxies studying the posterior distribution of an
impact parameter defined upon phase-space and size
information. The evolutionary status of HVS scattering candidates is further analyzed
based to their Hertzsprung--Russell (HR) diagram to confirm that their expected age is consistent with having travelled over typically large distances.

In section~\ref{sec:data} we illustrate our \textit{Gaia} DR2 HVS catalog, and the data selection for Milky Way GC and nearby
galaxies discussing distribution and kinematic properties properties
of each selection. In section~\ref{sec:meth} we outline our Bayesian
methodology to evaluate the scattering evidence between HVS and compact objects in GC or
nearby galaxies. In section~\ref{sec:results} we give our results. We
conclude in section~\ref{sec:conclusions}. In appendix~\ref{sec:altercuts} we discuss results based on an alternative stellar selection than the one considered throughout the main paper. In appendix~\ref{sec:coord}
we define our Galactocentric reference frame. In
appendix~\ref{sec:datarefs} we provide references for the selected GC
and galaxies, as well as orbit data for Sagittarius dSph and HVS
compatible with having crossed its trajectory.

\section{Data}
\label{sec:data}

\subsection{Hyper-velocity stars}
\label{sec:hvs}

Our reference catalog is \textit{Gaia} DR2  \citep{2016A&A...595A...1G, 2018A&A...616A...1G}, an all sky survey consiting of more than 1.3 billion stars. It contains accurate accurate positions ($\alpha$, $\delta$), proper motions ($\mu_{\alpha*}$,$\mu_{\delta}$), parallax ($\omega$), radial velocity, magnitudes and colors for the bright end, for $\sim7$ million stars.
We base our analysis on the 7183262 stars selection provided by~\citet{2018MNRAS.tmp.2466M}, established with the following quality cuts (see also \textit{Gaia} DR2 documentation\footnote{\url{https://gea.esac.esa.int/archive/documentation/GDR2/}} for more information about variables description):
\begin{itemize}
    \item \texttt{astrometric\_gof\_al} $< 3$.
    \item \texttt{astrometric\_excess\_noise\_sig} $\leq 2$.
    \item $-0.23 \leq$ \texttt{mean\_varpi\_factor\_al} $\leq 0.32$. 
    \item \texttt{visibility\_periods\_used} $> 8$.
    \item \texttt{rv\_nb\_transits} $> 5$.
\end{itemize}
We further clean the sample:
\begin{itemize}
    \item Selecting heliocentric total velocities in the Galactic rest frame large enough compared to their uncertainties $v-\sigma_v \gtrsim 500$~km/s, to mitigate errors in the Galactic absolute velocity.
    \item Removing potentially spurious radial velocities \citep{2019MNRAS.486.2618B}.\footnote{The list of possibly contaminated radial velocities is available as ancillary file to \href{http://arxiv.org/abs/1901.10460}{arXiv:1901.10460} [astro-ph.SR].}
\end{itemize}
\citet{2018MNRAS.tmp.2466M} provides for each star the probability $P_{\rm ub}$ of being unbound to the Milky Way potential. Our final selection of 1649 stars takes into account this information and further photometric and astrometric quality cuts \citep{2019MNRAS.487.3568S}:
\begin{itemize}
    \item $P_{\rm ub} > 0.5$.
    \item Color cut $G_{\rm BP} - G_{\rm RP} < 1.5$~mag.
    \item Magnitude cut $G < 14.5$~mag, and $G, G_{\rm BP}, G_{\rm RP} > 0$~mag.
    \item BP-RP excess flux factor cut $1.172 < E_{\rm bprp} < 1.3$.
    \item \texttt{astrometric\_excess\_noise} $< 1$.
\end{itemize}
See appendix~\ref{sec:altercuts} for an alternative stellar selection.

We consider distance estimates given in \citet{2018MNRAS.tmp.2466M}. The catalog does not include systematic errors on parallax measurements, and it sets a large scale length in the distance prior for likely distant stars that may overestimate distances. To overcome these shortcomings, we also consider the more recent \citet{2019A&A...628A..94A} work that improves the accuracy of extinction and effective temperature estimates provided with \textit{Gaia} DR2 by combining its astrometric and photometric measurements with external photometric catalogs. In particular, the related distance catalog includes errors induced by a parallax zero-point offset. We find that \citet{2018MNRAS.tmp.2466M} distances are systematically larger on average by roughly a factor of 2 than \citet{2019A&A...628A..94A} estimates for distant stars. The disagreement increases even up to an order of magnitude for stars whose distance is estimated to be $\lesssim 2$~kpc by \citet{2019A&A...628A..94A}, whereas this latter catalog agrees well with other computations \citep{2018AJ....156...58B, 2019MNRAS.487.3568S} within this range. 
Probabilities $P_{\rm ub}$ used for our selection rely on \citet{2018MNRAS.tmp.2466M} distances, but evaluating $P_{\rm ub}$ based on \citet{2019A&A...628A..94A} is beyond the scope of the present work. Since $P_{\rm ub}$ does not enter in following computations, comparing results based on the two different distance estimates still provides a consistency check.

\begin{figure}
  \centering
  \includegraphics[width=\columnwidth]{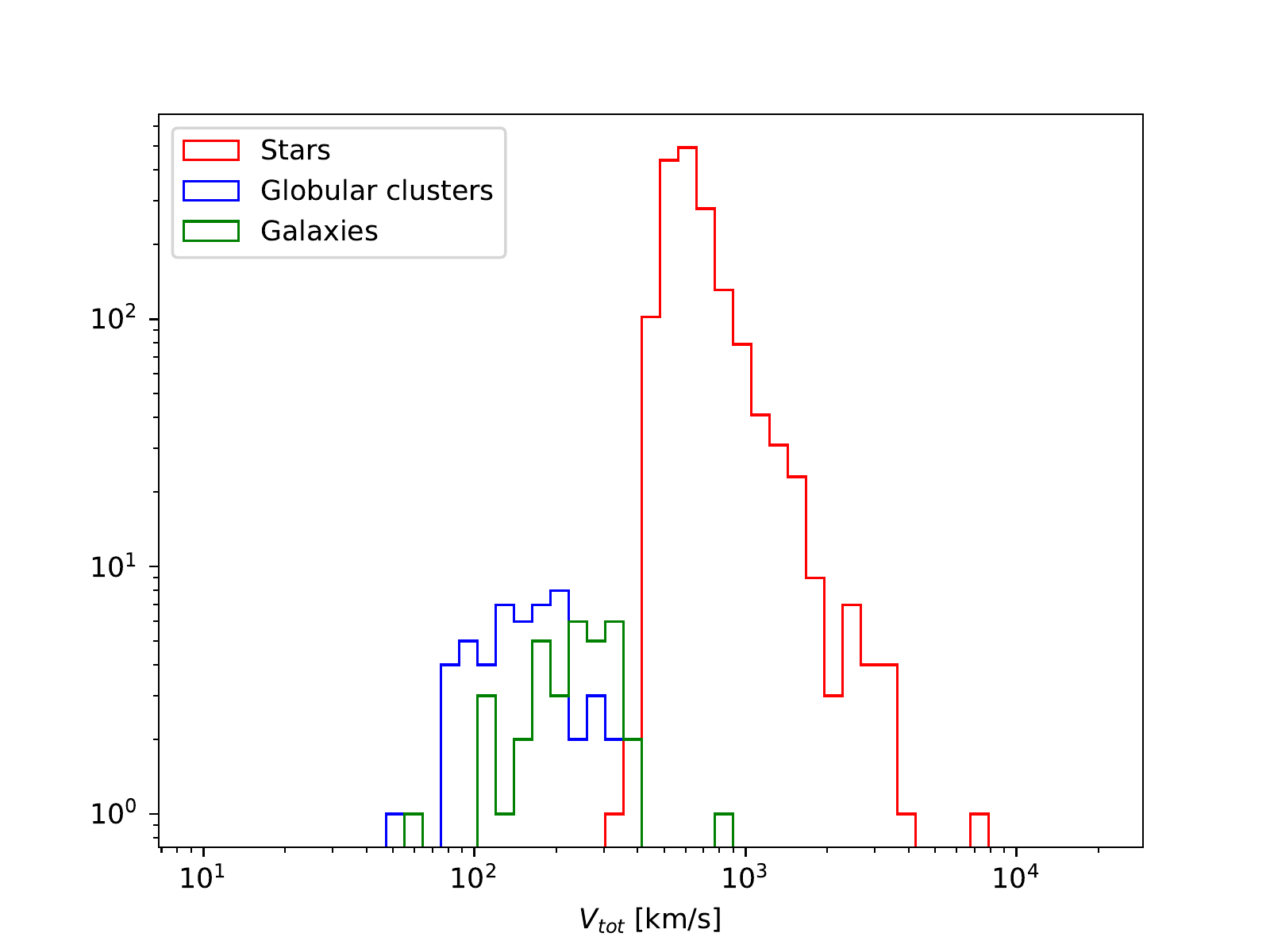}
  \caption{Total velocities in Galactocentric coordinates for our
    selection of stars, globular clusters and galaxies.}
  \label{fig:vel_hist}
\end{figure}

\begin{figure}
  \centering
  \includegraphics[width=\columnwidth]{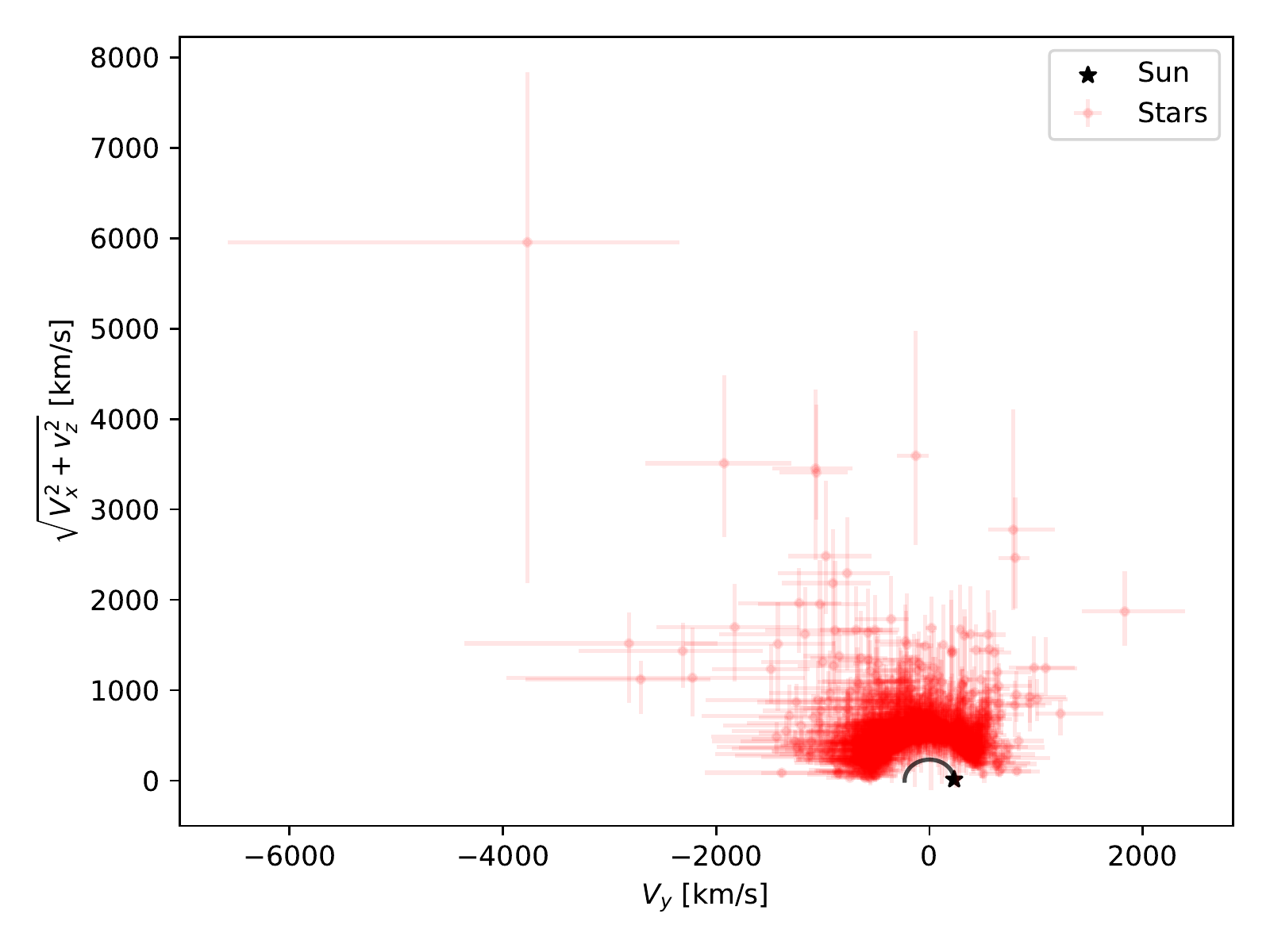}
  \caption{Toomre diagram in terms of Galactocentric Cartesian
    velocities for our HVS selection. All HVS are above the black
    semi-circle centered at the origin and of radius given by the Sun
    (black star) $V_y$ component.}
  \label{fig:toomre}
\end{figure}

\begin{figure}
  \centering
  \includegraphics[width=\columnwidth]{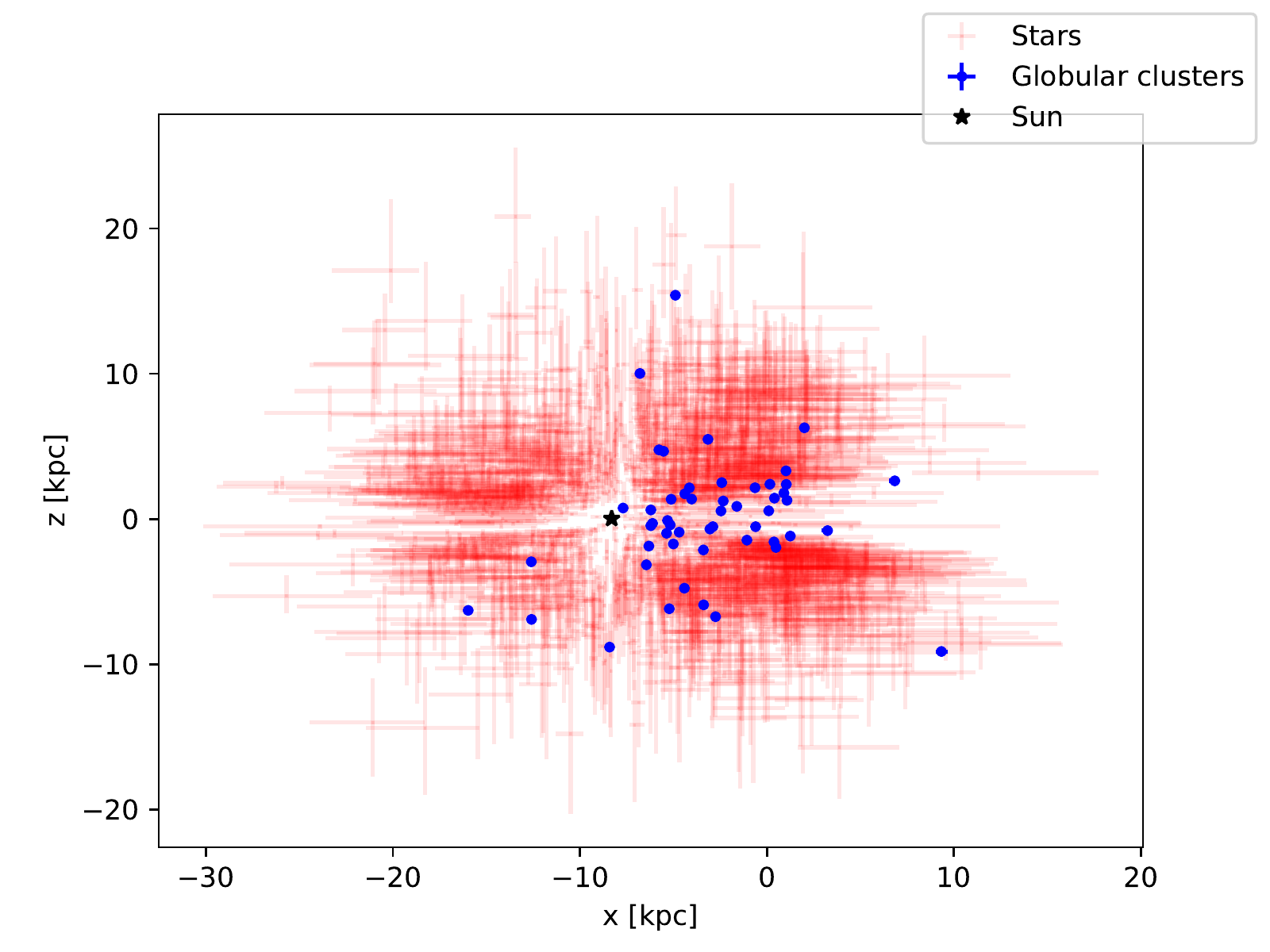}
  \includegraphics[width=\columnwidth]{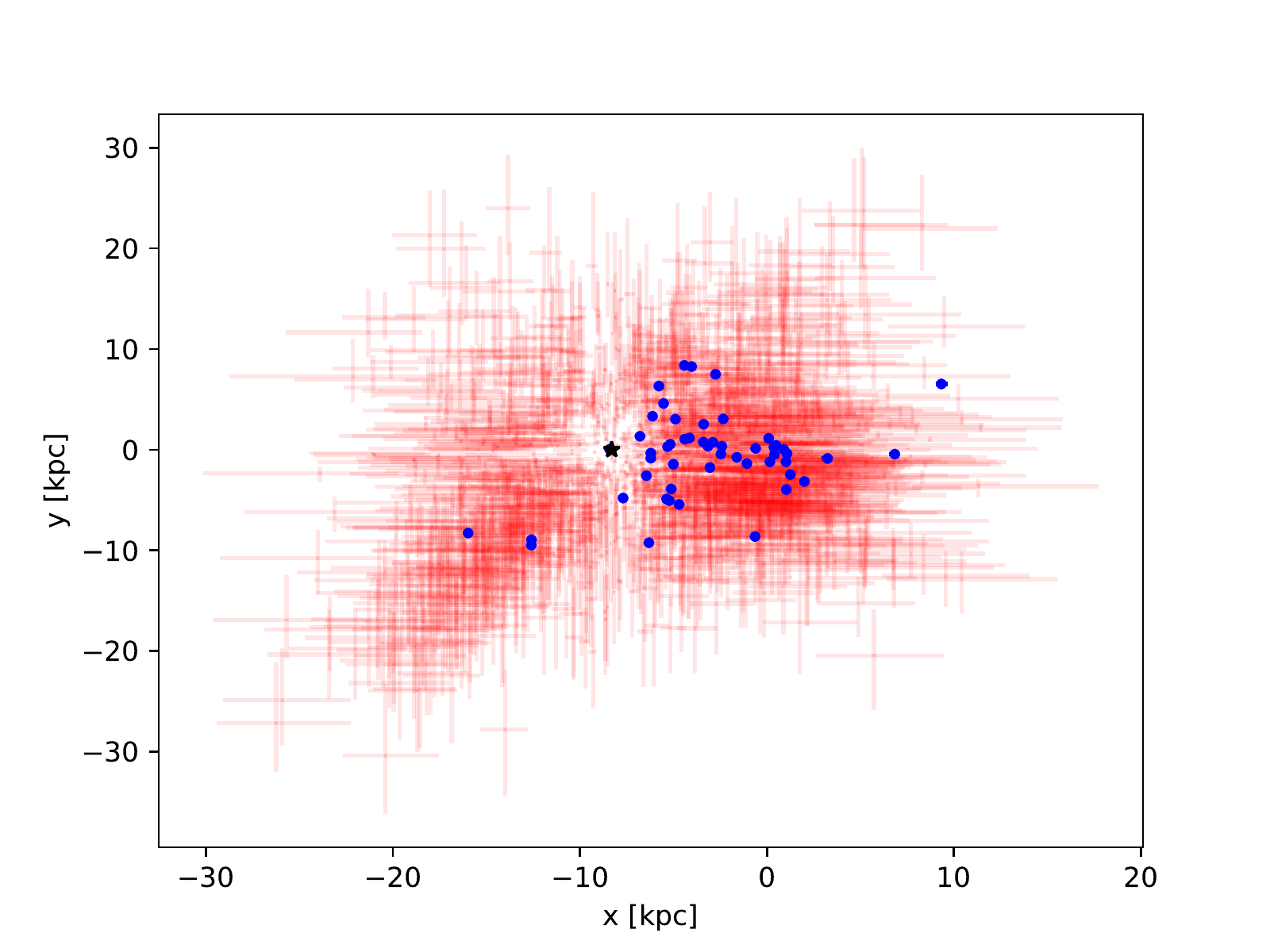}
  \caption{Galactocentric positions for stars and globular
    clusters. Error bars $\lesssim 3\%$ for GC are smaller
    than markers (the latter are not representative of objects
    extension).}
  \label{fig:pos_gaia}
\end{figure}

Figure \ref{fig:vel_hist} shows that most of our HVS selection is characterized by total Galactocentric velocities
$V_{tot} \sim \mathcal{O}(10^3)$~km/s. For details about our coordinates system see
appendix~\ref{sec:coord}. Figure \ref{fig:toomre} shows the Toomre
diagram with the Galactocentric Cartesian $V_y$ component on the
abscissa, and the $\sqrt{V_x^2 + V_z^2}$ component on the
ordinate.\footnote{The Toomre diagram is often expressed in terms of
  Galactic (heliocentric) Cartesian velocities U, V, W
  \citep[e.g.,][]{2012MNRAS.427..274S} well suited to describe the
  solar neighborhood. Here we are interested in the dynamics of the
  Galaxy on a global scale and a Galactocentric frame is more
  convenient.}  Here and afterwards error bars indicates 68\%
confidence intervals. The diagram is populated only above a semi-circle centered at
the origin and of radius given by the Sun $V_y$ component, suggesting
that we select a population of halo stars
\citep[e.g.,][]{2017ApJ...845..101B}. Figure \ref{fig:pos_gaia} shows
Galactocentric positions with error bars dominated by uncertainties on
distances from the Sun.

\subsection{Globular clusters and galaxies}
\label{sec:gc_gal}

We select 52 globular clusters (GC) identified in the \textit{Gaia} DR2 catalog
\citep{2018A&A...616A..12G} that report both full phase-space and
radius, defined as the maximum radius at which proper-motion members
are found.

We select all galaxies for which we can match position and half-light
radius (measured along major axes) given by\footnote{We use the table updated on 20
  September 2015 available at
  \url{http://www.astro.uvic.ca/~alan/Nearby_Dwarf_Database.html}.}
\citet{2012AJ....144....4M}
with peculiar motions of identified dwarf Milky Way satellite galaxies in
\textit{Gaia} DR2 \citep{2018A&A...619A.103F}. We also include Antlia II
\citep{Torrealba:2018fwy} and Andromeda \citep[velocity
from][]{2012ApJ...753....8V}.\footnote{\citet{2012ApJ...753....8V}
  assumes $V_y = 239 \pm 5$~km/s for the Sun Galactocentric velocity
  component along the direction of Galactic rotation, while here we
  assume 232.24~km/s (see appendix~\ref{sec:coord}). The difference is
  negligible for our purposes.} In the latter case we retrieve the
optical major-axis from the SIMBAD database \citep{Wenger:2000sw},
also used to get sizes and velocities of the large (LMC) and small
(SMC) Magellanic clouds that we combine with positions listed in
\citet{2012AJ....144....4M}. This gives a total of 35 galaxies. 

\begin{figure}
  \centering
  \includegraphics[width=\columnwidth]{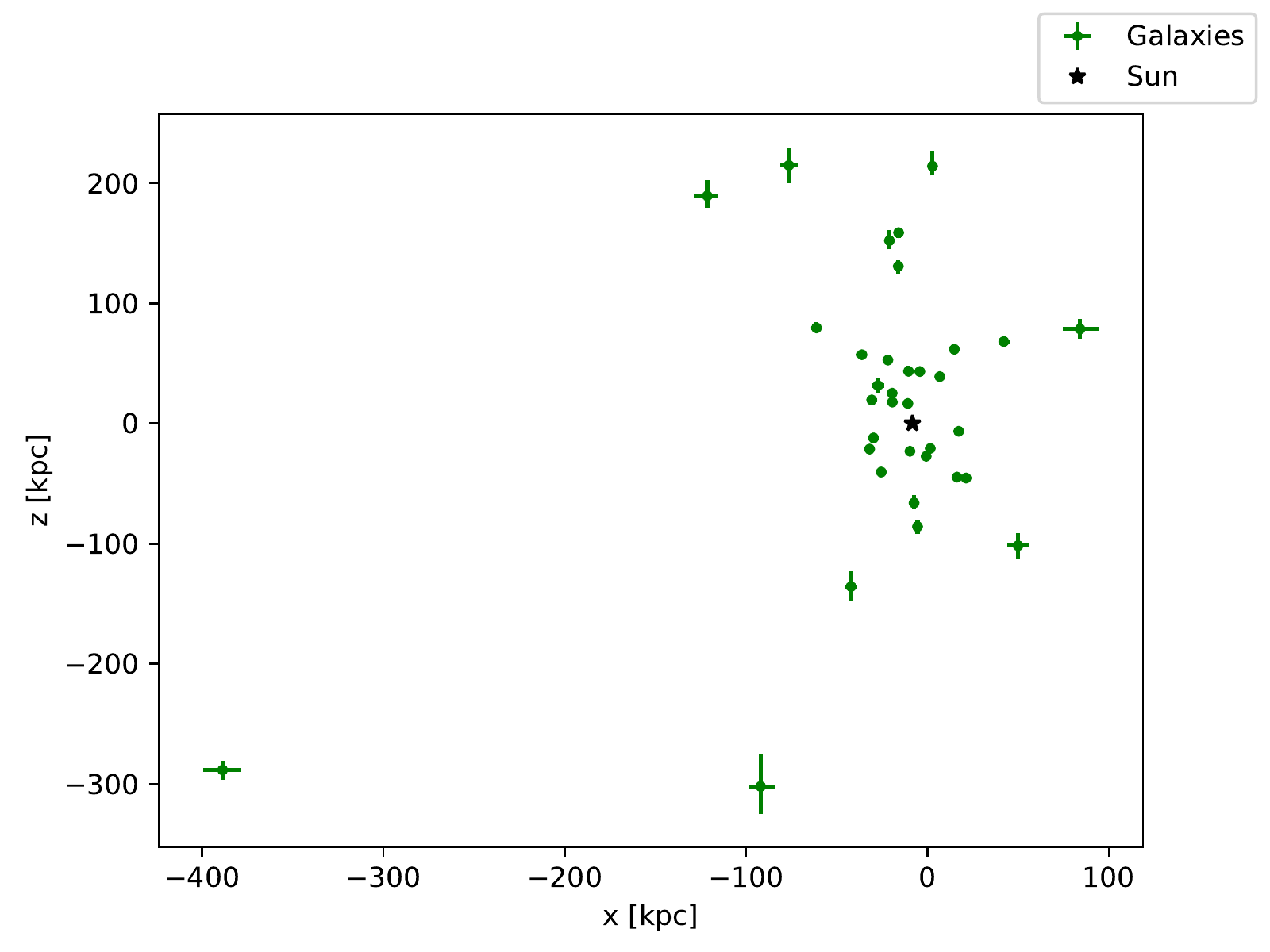}
  \includegraphics[width=\columnwidth]{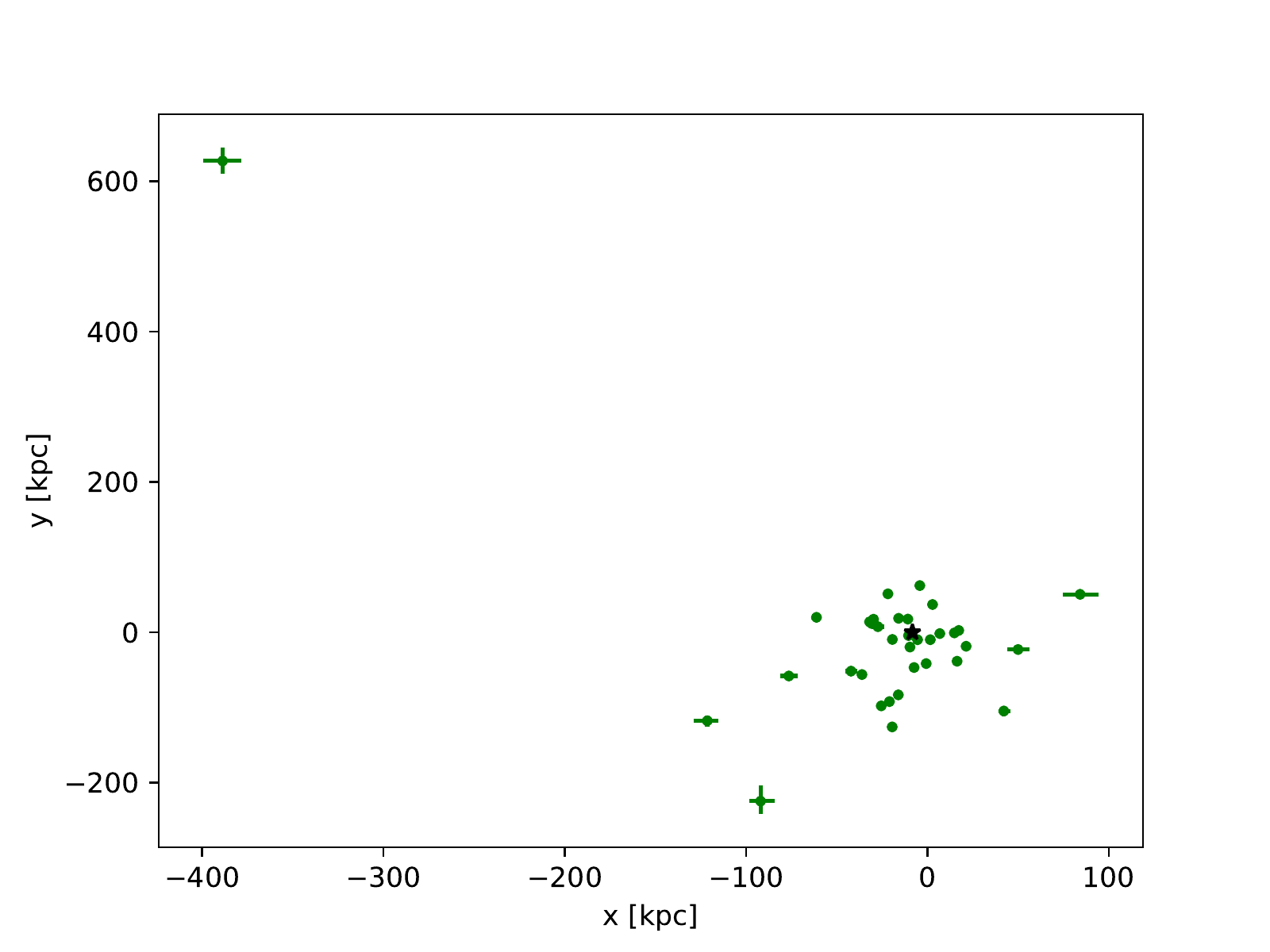}
  \caption{Galactocentric positions for galaxies. The further galaxy
    is Andromeda. Marker size (for some object larger that error bars)
    is not representative of object extension.}
  \label{fig:pos_gal}
\end{figure}

\begin{figure}
  \centering
  \includegraphics[width=\columnwidth]{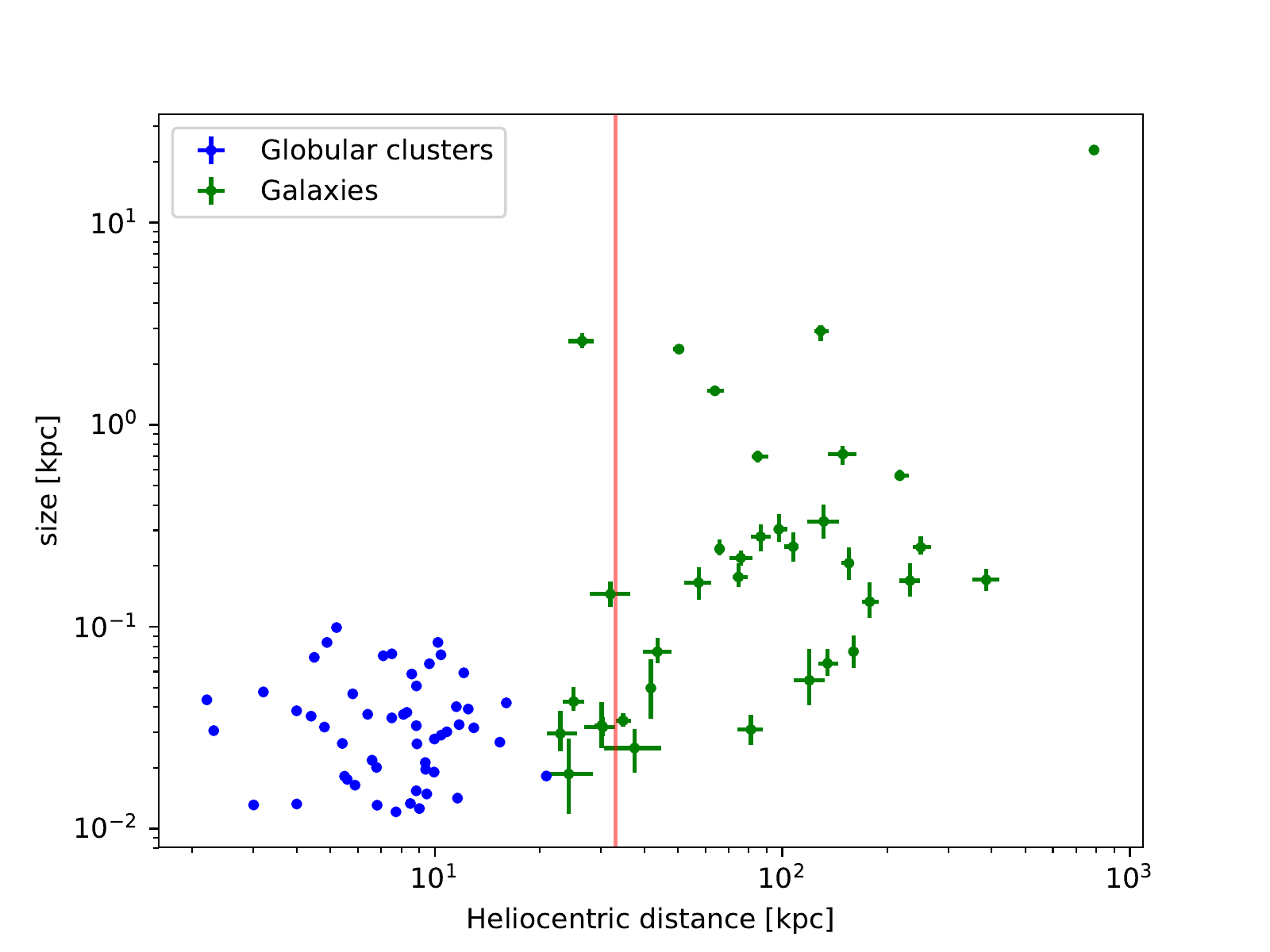}
  \caption{Heliocentric distance and size for GC and galaxies. The
    vertical line corresponds to the further HVS; the largest
    object within this distance is Sagittarius dSph, placed at
    coordinates $(26 \pm 2, 2.6 \pm 0.2)$~kpc.}
  \label{fig:size}
\end{figure}

Figure \ref{fig:vel_hist} shows total Galactocentric velocities
$V_{tot} \sim \mathcal{O}(10^2)$~km/s both for GC and
galaxies. Figures \ref{fig:pos_gaia} and \ref{fig:pos_gal} show GC and
galaxy positions, respectively. We verified that the asymmetric GC
distribution, concentrated between the Sun and the galactic Center is
also present in the full catalog and not only in our selection (note
the lack of GC in the upper quadrants for objects with $x<x_{\sun}\approx-8.3$~kpc).  Figure
\ref{fig:size} shows heliocentric distances and sizes in terms of radii
defined above. GC are characterized by sizes $\mathcal{O}(10)$ --
$\mathcal{O}(100)$~pc, while galaxies show a larger variation because
we include objects of different types, ranging from dwarfs with radius
$\sim\mathcal{O}(10)$~pc to large galaxies such as Andromeda with
major semi-axis $\sim 23$~kpc. Most galaxies reach distances further
than our HVS selection.

Further details about the selected objects are given in
appendix~\ref{sec:datarefs}.

\section{Methodology}
\label{sec:meth}
\subsection{Trajectories}

Orbits are integrated using the \texttt{gala} software library
\citep{2017JOSS....2..388P}\footnote{\url{https://gala-astro.readthedocs.io}}.
We use its default Milky Way potential model based on four
components. Below $G$ indicates the gravitational constant, $r_s$ is
the radial distance in Galactocentric spherical coordinates, $r_c$ and
$z_c$ are the radial distance and height, respectively, in
Galactocentric cylindrical coordinates. The nucleus and bulge follow a
Hernquist potential for a spheroid \citep{1990ApJ...356..359H}:
\begin{equation}
  \phi(r_s) = -\frac{Gm}{r_s + a} \;.
\end{equation}
The disk follows a Miyamoto-Nagai profile \citep{1975PASJ...27..533M,
  2015ApJS..216...29B}:
\begin{equation}
  \phi(r_c, z_c) = -\frac{Gm}{\sqrt{r_c^2 + \left(a + \sqrt{z_c^2 + b^2} \right)^2}} \;.
\end{equation}
The halo follows a spherical Navarro-Frenk-White (NFW) profile:
\citep{Navarro:1995iw}.
\begin{equation}
  \phi(r_s) = -\frac{Gm}{r_s} \ln\left( 1 + \frac{r_s}{a} \right) \;.
\end{equation}
We use the same parameters as in \citet{2018MNRAS.tmp.2466M},
summarized in table \ref{tab:mwpot}. In the PBH CDM scenario the inner
part of the halo is better described by an Einasto profile
\citep{Einasto:1965czb}, but this only affects the dynamics close to
the Galactic center, where a few of our HVS candidates lie. We have
verified that taking into account a power-law gravitational
potential profile~\citep{1993MNRAS.260..191E, 1994MNRAS.267..333E, Calcino:2018mwh} rather than NFW does not affect our
conclusions.

\begin{table}
  \centering
  \caption{Milky Way potential parameters.}
  \label{tab:mwpot}
  \begin{tabular}{lll}
    \hline
    Component & Potential      & Parameters \\
    \hline
    Nucleus   & Hernquist      & $m=1.71 \times 10^9 M_{\sun}$ \\
              &                & $a=0.07$~kpc \\
    Bulge     & Hernquist      & $m=5 \times 10^9 M_{\sun}$ \\
              &                & $a=1.0$~kpc \\
    Disk      & Miyamoto-Nagai & $m=6.8 \times 10^{10} M_{\sun}$ \\
              &                & $a=3$~kpc \\
              &                & $b=0.28$~kpc \\
    Halo      & NFW            & $m=5.4 \times 10^{11} M_{\sun}$ \\
              &                & $a=15.62$~kpc \\
\hline
  \end{tabular}
\end{table}

We trace trajectories back in time by 100~Myr and 1~Gyr when looking
for correlations between HVS and GC or galaxies, respectively, with a
resolution of 1~kyr (necessary to resolve the smallest GC). In the
case of large and distant objects (Andromeda, LMC and SMC) we use a
poorer time resolution of 100~kyr, but integrate up to 5~Gyr back in
time (to assure that the slower stars in our selection have the time
to reach Andromeda distance).\footnote{More precisely, we integrate
  orbits setting time steps of 0.1~Myr (10~Myr for Andromeda, LMC and
  SMC) in \texttt{gala} and then interpolate with cubic splines to reach the desired time
  resolution.}

\subsection{Impact parameter}
\label{sec:theta}

Let $r(t_i)$ be the distance between a star and a GC or galaxy with
radius $R$ (defined as discussed in section \ref{sec:data}) at a given
time $t_i$. We define the impact parameter for the trajectories of a
star and a GC/galaxy as
\begin{equation}
\theta \equiv \frac{1}{R} \min_i r(t_i) \;.
\end{equation}

Given the impact parameter likelihood $P(D|\theta)$, where $D$ denotes
data, we want to identify those stars compatible with having scattered
with compact objects bounded to a given GC or galaxy. In the case of GC the radius
is determined via proper motion members, and we set $\theta \lesssim 1$ as necessary condition for scattering. In the case of galaxies, the
half-light radius (measured along the major axis) or the optical major
semi-axis are looser proxies of the underlying DM
distribution. Furthermore, dwarf galaxies can have large
ellipticity. We take into account these uncertainties by extending the
relevant range to $\theta \lesssim 10$ for a star being compatible
with having scattered with compact objects bound to a given galaxy, and by studying results as functions of the impact parameter.

\subsection{Likelihoods}
\label{sec:modfit}

We illustrate how we sample from trajectory parameters space based on
observables or derived quantities available in the catalogs described
in section \ref{sec:data}. Since position errors are dominated by
uncertainties on distances, we neglect errors in right-ascension
$\alpha$ and declination $\delta$ for computational convenience.

We write the probability distribution for the star trajectory
parameters in terms of log-normal\footnote{The ${\rm Lognormal}(m_x, v_x)$ distribution parameters are related to the mean value $x$ and variance $\sigma_x^2$ of the random variable by $m_x = \ln\left(x / \sqrt{1+\sigma_x^2/x^2}\right)$ and $v_x = \ln\left(1+\sigma_x^2/x^2\right)$.} and normal $\mathcal{N}(\mu, \sigma^2)$
distributions\footnote{Some of the data discussed here provides 16\%
  and 84\% quantiles rather than the variance. We approximate the
  log-normal or normal distributions variance as the mean of these
  lower and upper bounds.}
\begin{eqnarray}
  P_* &=&
          {\rm Lognormal}\left(m_d, v_d\right)
          \mathcal{N}(\boldsymbol{\mu}_{\alpha\beta}, \boldsymbol{\Sigma}_{\mu_{\alpha\beta}})
         \mathcal{N}(V_r, \sigma_{V_r}^2) \;.
\end{eqnarray}
Here $d$ is the heliocentric distance discussed in section~\ref{sec:hvs} (we
find that a log-normal distribution for distances recover the respective
asymmetric probability distributions), $\boldsymbol{\mu}_{\alpha\beta} = \left(\mu_{\alpha^*}, \mu_{\delta}\right)$ with
$\boldsymbol{\Sigma}_{\mu_{\alpha\beta}}$ their covariance matrix ($\mu_{\alpha^*}=\mu_{\alpha}cos(\delta)$ is
proportional to the proper motion in right-ascension direction
$\mu_{\alpha}$, $\mu_{\delta}$ is the proper motion in declination
direction), and $V_r$ is the radial velocity. \textit{Gaia} DR2 provides astrometric parameters at epoch J2015.5 that, for comparison with other datasets,
we transform to epoch J2000.0 following the reduction procedures used to construct the
Hipparcos and Tycho catalogues \citep{1997ESASP1200.....E}\footnote{We use epoch propagation functions provided by TOPCAT \citep{2005ASPC..347...29T}.} (more
rigorous transformations including the effects of light-travel time
are given in \citet{2014A&A...570A..62B}, but they are not well suited
for negative parallaxes characterizing some of our sources). We verified that the only quantities affected non-negligibly by epoch propagation are $\alpha$, $\delta$ and for a few sources $V_r$.

We write the probability distribution for globular clusters as
\begin{equation}
  P_{gc} =
  \mathcal{N}\left(\boldsymbol{X}, \boldsymbol{\Sigma_X}\right)
  \mathcal{N}(\boldsymbol{\mu}_{\alpha\beta}, \boldsymbol{\Sigma}_{\mu_{\alpha\beta}})
  \mathcal{N}(V_r, \sigma_{V_r}^2) \;,
\end{equation}
where $\boldsymbol{X}$ are heliocentric Cartesian coordinates used to
compute distances. In the case
of galaxies we have
\begin{eqnarray}
  P_{gal} &=&
  \mathcal{N}\left(\mu, \sigma_\mu^2\right)
  \mathcal{N}(\boldsymbol{\mu}_{\alpha\beta}, \boldsymbol{\Sigma}_{\mu_{\alpha\beta}})
  \mathcal{N}(V_r, \sigma_{V_r}^2) \times
  \nonumber \\
  &&{\rm Lognormal}(m_R, v_R) \;.
\end{eqnarray}
Here the heliocentric distance is computed from the distance modulus $\mu$
(for Antlia II we use the derived distance obtained in
\citet{Torrealba:2018fwy}). For Andromeda we sample directly from the
derived Galactocentric velocity error distributions
\citep{2012ApJ...753....8V} rather than proper motions and radial
velocity. The log-normal distribution for the radius $R$ takes into
account the asymmetric error bounds for some of the objects, and the
fact that its expectation value is restricted to be positive (we
verified that results do not change if we assume a normal distribution
and a prior $R>0$). In the case of Andromeda, LMC, SMC and of GC we
don't have information about the respective radius probability
distributions, but this is not critical information since afterwards
we study results as a function of the impact parameter
$\theta \propto 1/R$.

The impact parameter likelihood $P(D | \theta)$ for every star and GC
or galaxy pair is obtained by sampling from $P_*P_{gc}$ or $P_*P_{gal}$,
respectively. We reconstruct each likelihood drawing at least 1000
random samples, necessary to recover Bayes factors at the
$\mathcal{O}(1\%)$ level.\footnote{While the sampling can be in
  principle parallelized over each star and GC/galaxy pair, we are
  limited by high memory costs to reconstruct only a few likelihoods
  at the time. This prevents us from running a Markov chain Monte
  Carlo sampler for each case.} We verified that uncertainties in the
Galactocentric frame definition, see appendix~\ref{sec:coord}, are
negligible for our purposes.\footnote{For consistency we point out that for the cases shown in
  section \ref{sec:results} figures we do sample also from uncertainties in the
Galactocentric frame.}

We verified that for the cases of our interest ($\theta \lesssim 10$)
likelihoods are well fit by skew log-normal
distributions\footnote{While this form is not well suited for
  $\theta \to 0$, we find it reliable down to our smallest sampled
  values $\theta \sim \mathcal{O}(10^{-2})$.}
\begin{equation} \label{eq:logfit}
  P(D | \theta)
  = \frac{1}{(\theta - \lambda_D)\sigma_D\sqrt{2\pi}}
    \exp\left[ -\frac{\left(\ln(\theta-\lambda_D) - \mu_D\right)^2}{2\sigma_D^2}
    \right] \;,
\end{equation}
where the fit parameters $\mu_D$, $\sigma_D$, $\lambda_D$ are
determined for each star and galaxy/GC pair.

Together with the impact parameter likelihoods, we also reconstruct
the likelihoods for the scattering time (i.e., the time corresponding
to the minimum distance between trajectories).

\subsection{Hypothesis testing}

Given an impact parameter likelihood and a prior $\Pi(\theta)$, the
posterior distribution is given by Bayes' theorem
$P(\theta | D) \propto P(D | \theta) \Pi(\theta)$. We can establish at
what credible interval a star is compatible with having scattered with
compact objects bound to the given system if values $\theta \lesssim \mathcal{O}(1)$ (for GC)
or $\theta \lesssim \mathcal{O}(10)$ (galaxies) are included in the
region of interest. Below we consider Bayesian hypothesis testing.

We want to compute the Bayes factor for the hypothesis $H$ that the
star trajectory intersects the given galaxy/GC trajectory, relative to
the hypothesis $\bar{H}$ of no intersection. Positive evidence for $H$
suggests that the star is compatible with having scattered with compact objects
bounded to the given galaxy/GC (possibly fixing a lower threshold for
$\theta$).

Given the likelihood $P(D | \theta)$ computed in section
\ref{sec:modfit}, the marginal likelihoods under the two hypothesis
are:
\begin{eqnarray}
  P(D | H) &=& \int_0^\infty \dd\theta\ P(D | \theta) \Pi(\theta | H) \;,
  \\
  P(D | \bar{H}) &=& 1 - P(D | H) \;.
\end{eqnarray}
$\Pi(\theta | H)$ models our prior knowledge on the impact parameter
distribution under the trajectories intersection hypothesis. We opt for a flat prior
\begin{equation}
  \Pi(\theta | H) =
  \begin{cases}
    \frac{1}{\theta_* - \theta_0} & \theta_0 \leq \theta \leq \theta_* \\
    0 & \text{otherwise.}
  \end{cases}
\end{equation}
Finally, the Bayes factor is defined by:
\begin{equation}
  K(\theta_0, \theta_*) \equiv \frac{P(D | H)}{P(D | \bar{H})} \;.
\end{equation}

We fix the lower prior threshold $\theta_0$ at the smallest distance from the GC/galaxy center at which we expect scattering with compact objects. Note that in the PBH CDM scenario, due to the finite size of black holes, the DM density distribution, rather than peaking around a central cusp, may be small in the innermost regions of the galaxy due to the gravitational slingshot effect~\citep{Garcia-Bellido:2017fdg}, and reach a maximum at a finite distance from the center. We then
study the Bayes factor as a function of the upper threshold
$\theta_*$. In other words, we marginalize over the uncertainties
outlined in section \ref{sec:theta}.

Our prior choice is dictated by simplicity given that here we aim at investigating at once several objects with different mass distributions (in some case highly irregular) dependent not only on astrophysical details associated to a given object, but also on the intrinsic nature of DM. Follow-up analyses focused on individual objects should consider priors based on realistic modeling for ejection location and velocity inside a GC/galaxy. We stress that being the problem inherently Bayesian (we have at our disposal only one physical realization of the sources under consideration), discussion cannot disregard a prior choice.

Given the flat prior, the marginal likelihood can be written in terms
of the likelihood cumulative distribution function (CDF). While we
could compute the empirical CDF, it is numerically convenient to use
the analytical form obtained by fitting eq.~(\ref{eq:logfit}), and
obtain an analytical expression for the marginal likelihood
\begin{eqnarray} \label{eq:margL_fit}
  P(D | H) &=& \frac{F(\theta_*) - F(\theta_0)}{\theta_*-\theta_0}\;,
\end{eqnarray}
where $F(x)$ is the CDF up to a given threshold $x$, given in
terms of the CDF for the standard normal distribution $\Phi$
\begin{eqnarray}
  F(x) &=&  \Phi\left( \frac{\ln(x-\lambda_D)
           -\mu_D}{\sigma_D} \right)
           \nonumber \\
       &=& \frac{1}{2} \left[ 1 + {\rm erf}\left(
           \frac{\ln(x-\lambda_D)
           -\mu_D}{\sigma_D\sqrt{2}}  \right)  \right]\;,
\end{eqnarray}
and erf is the error function. We
verified that differences in Bayes factors computed using the
empirical CDF and the analytical approximation are of order 10\% for
likelihoods peaked in the $\theta$ range of interest, and within 5\%
for those objects with $K \gtrsim 1$.

\section{Results}
\label{sec:results}

\begin{figure}
  \centering
  \includegraphics[width=\columnwidth]{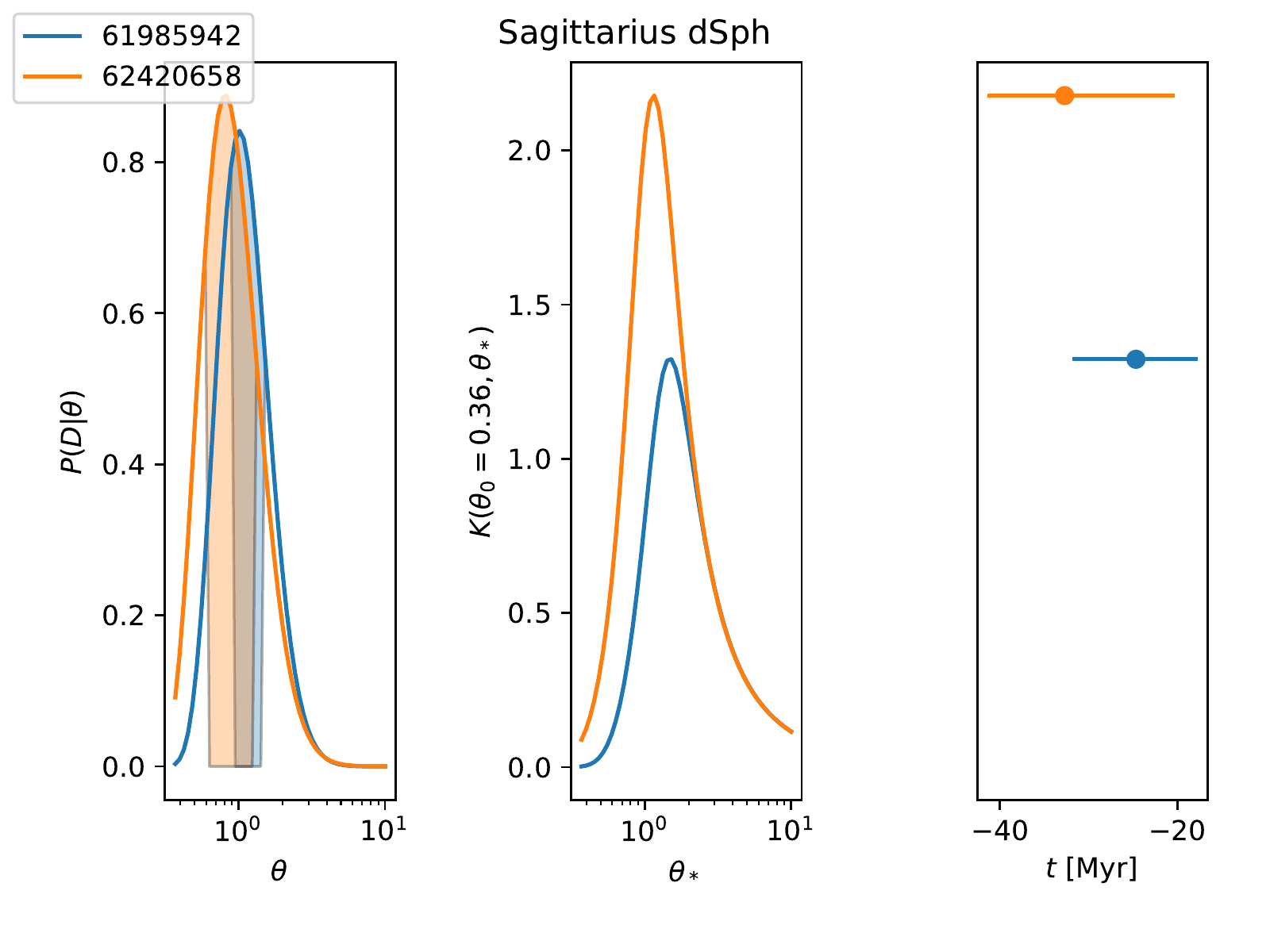}
  \caption{Stars compatible with having scattered with compact objects within Sagittarius
    dSph assuming \citet{2018MNRAS.tmp.2466M} distances. The legend shows the first 8 digits of \textit{Gaia} DR2 star
    identification numbers. \emph{Left panel:} Impact parameter
    likelihood. Filled regions are 68\% confidence
    intervals. \emph{Central panel:} Bayes factor for $\theta_0=0.36$ as
    a function of the upper prior threshold $\theta_*$. \emph{Right
      panel:} Mean and 68\% intervals for the intersection time (today corresponds to $t=0$~yr). The
    height of error bars is proportional to the respective maximum
    Bayes factor.}
  \label{fig:sag}
\end{figure}

We first fix the prior lower threshold $\theta_0=0$ and search for
scattering events
within a given upper threshold $\theta_*$. Then we repeat the search for a few values of the lower prior
threshold $\theta_0 \leq 0.5$. This parameterise our belief that a
scattering event is unlikely to take place in the innermost region of
a given object, as it is the case if we look for interactions with the
PBH CDM halo of a galaxy. We consider both \citet{2018MNRAS.tmp.2466M} and \citet{2019A&A...628A..94A} distance estimates. If $K>1$ then the star is
compatible with having interacted with a given object.

In the case $\theta_0=0$ we do not find candidate scattering events (see however appendix~\ref{sec:altercuts}). As shown in figure~\ref{fig:sag}, in the case of the
Sagittarius dwarf spheroidal galaxy (Sagittarius dSph), we find two stars candidate to having scattered with compact objects within it when assuming \citet{2018MNRAS.tmp.2466M} distances for $\theta_0=0.36$, corresponding to the
minor-to-major axis ratio for Sagittarius dSph given its ellipticity
$\epsilon=0.64\pm0.02$ \citep{2012AJ....144....4M}. The
Bayes factor peaks at values $\theta_* \sim 1$. Scattering times are
about 20--40~Myr ago, excluding that these events are directly related
to the fact that Sagittarius dSph may have crossed the Galactic disk
300--900~Myr ago \citep{2018Natur.561..360A}.

\begin{figure}
  \centering
  \includegraphics[width=\columnwidth]{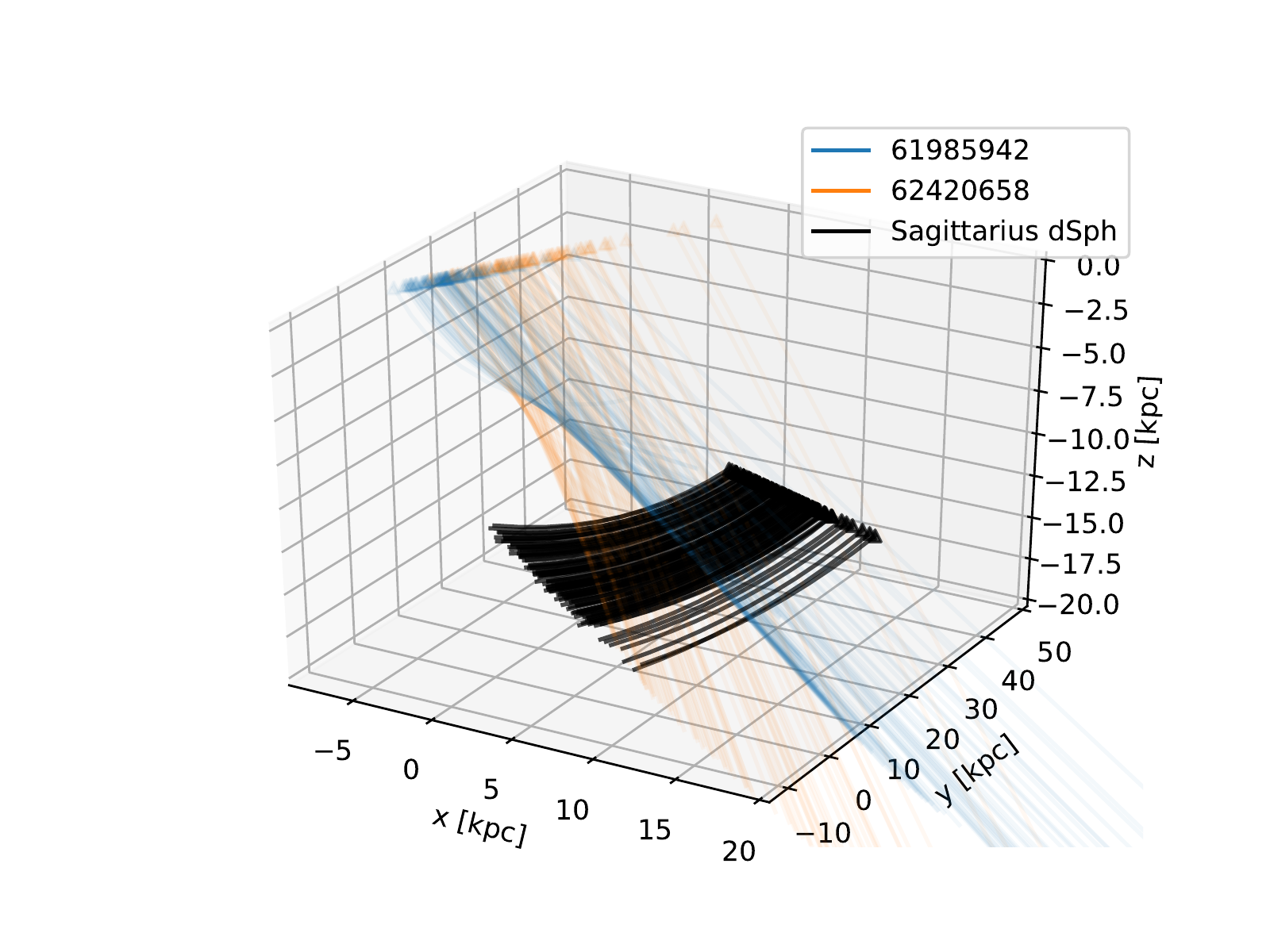}
  \caption{Orbits in Galactocentric Cartesian coordinates
    corresponding the same objects discussed in
    figure~\ref{fig:sag}. Line density is
    proportional to the likelihood. Markers represent positions
    today.}
  \label{fig:orbit}
\end{figure}

Figure~\ref{fig:orbit} shows 100 orbits randomly sampled from the
respective likelihoods for objects discussed in
figure~\ref{fig:sag}. Distance uncertainties
lead to large spreads in radial directions from the Sun. Trajectories
beyond the scattering event are no longer reliable as they may be
depend on a different potential, but that's not a concern for our
purposes.

In order to verify whether our results are compatible with the evolutionary status of our stars (i.e., whether their life time is enough to have crossed the very large distance from the original point), we derive their effective temperature (T$_{\rm eff}$) and bolometric luminosity (L$_{\rm bol}$) and compare with theoretical isochrones and evolutionary tracks. First, we obtain L$_{\rm bol}$ and T$_{\rm eff}$ using the Virtual Observatory SED Analyzer \citep[VOSA,][]{Bayo2008-VOSA}, by constructing a complete Spectral Energy Distribution (SED) taking advantage of different photometric repositories under Virtual Observatory (VO) protocols. Then, the basic properties are derived using atmospheric models by Kurucz \citep{Castelli1997-Kurucz}.  We impose restrictions on the surface gravity ($\log g=2$) and metallicity ([Fe/H]=-0.5), based on the expectations about these candidates (they should be giant or subgiant stars with low metallicity, since Sagittarius dSph has [Fe/H]$=-0.40\pm0.2$~dex \citep{2012AJ....144....4M}). In any case, SED fitting depends weakly on $\log g$ and our results are very similar if a larger degree of freedom is allowed. Thus, the T$_{\rm eff}$ range between 6000 and 4250~K, whereas the L$_{\rm bol}$ are bracketed between 620 and 2000~L$_\odot$. We have compared these values with PADOVA models \citep{Marigo2017-Padova}. Their position in a HR diagram clearly shows that they have masses in the range 3.5-6 M$\odot$ and they are either in the subgiant branch, close to the subgiant branch,  or the Blue loop. Therefore, we can establish a lower limit for the age, about 100~Myr, fully compatible with our expectations.

Positive evidence is not confirmed when using \citet{2019A&A...628A..94A} distances. However, our result provides motivation to model a prior based on the actual three-dimensional DM distribution of Sagittarius dSph, a difficult issue that has to take into account strong tidal disruption.

More information about the sources discussed here is given in table~\ref{tab:hvs}.

\section{Conclusions}
\label{sec:conclusions}

We have defined a Bayesian framework to search for correlations between 1642 \textit{Gaia} DR2 high-velocity star trajectories and those of 52 globular clusters (identified by \textit{Gaia} DR2) and 35 Milky Way dwarf and nearby galaxies. We report 2 stars candidate to have scattered with compact objects within Sagittarius dSph roughly between 20 and 40~Myr ago when assuming distances estimated in \citet{2018MNRAS.tmp.2466M}. Analysis of their evolutionary status leads to a lower bound of about 100~Myr for their age, fully compatible with the scattering time window. Results are not confirmed when assuming distances estimated in \citet{2019A&A...628A..94A}.

These events may correspond to DM scattering, if the latter is composed by PBH able to accelerate significantly a star upon encounter. In principle the reported number of scattering events may be used to validate this hypothesis. However, given the statistically small sample considered here and uncertainties in PBH mass distribution we cannot put meaningful limits.

\citet{2018MNRAS.tmp.2466M} reported HVS candidates compatible with extragalactic origin. Their trajectory may also be explained if they scattered with DM bounded to a galaxy or GC. We repeated our search including their final HVS selection of 20 stars (with probability of being unbound larger than 80\%), and we do not find evidence to support this hypothesis (furthermore several stars in their selection are contaminated by spurious radial velocities \citep{2019MNRAS.486.2618B}).

In defining the impact parameter based on the radius of a sphere centered around a given galaxy or GC, we have taken into account the necessity to analyze at once several heterogeneous objects. Studying the Bayes factor as a function of the marginal likelihood prior takes into account that the radius is only a proxy to the actual shape (e.g., Sagittarius dSph is characterized by a large ellipticity) or DM distribution (that can extend well beyond the optical size). Nevertheless, a follow-up analysis focused on Sagittarius dSph should define the impact parameter based on its actual shape and DM distribution, taking into account its time evolution.

Our results are highly dependent on the distance computation methodology. We find that \citet{2018MNRAS.tmp.2466M} distances are usually significantly larger than other catalogs for our stellar selection (see table~\ref{tab:hvs}). Distances estimated in \citet{2019A&A...628A..94A} are more  in agreement with other computations (e.g., \citep{2018AJ....156...58B} and \citet{2019MNRAS.487.3568S}) although results generally differ significantly for stars more distant than 3~kpc. This sensitivity on systematic effects (especially a global parallax zero-point difficult to model) calls for future confirmations based on more accurate parallax estimates.

\citet{2018MNRAS.476.4697M} showed that the majority of HVS expected to be detected by \textit{Gaia} are fainter than the limiting magnitude to obtain radial velocities in DR2. Future data releases will include a larger number of stars with full phase-space information, together with improved astronomy and photometry. As our analysis shows, the understanding of systematic errors up to distances of around $10$~kpc is crucial for a robust search. More accurate data will also help to avoid spurious HVS identification possibly affecting our selection.

The fact that we only find candidate scattering events within Sagittarius dSph, relatively large and close, may be due to a selection effect that can be included in future analyses. In fact, if the events under consideration correspond to DM scattering, then we would expect to be able to detect a similar number of interactions also with other large galaxies such as LMC. It is important to repeat and possibly extend the search when future \textit{Gaia} data releases will be available.

The main motivation of the present search is that HVS may be correlated to past scattering events with massive compact objects in dwarf galaxies. The same idea prompts another interesting prospect, i.e., using trajectories of HVS compatible with having extragalactic origin as a guide for discovery of faint dwarf galaxies, particularly relevant to extend catalogs of low surface brightness galaxies \citep{2019MNRAS.483.1754D} and for missions like the MESSIER surveyor \citep{2017IAUS..321..199V}.

\section*{Acknowledgements}

We thank Alex Drlica-Wagner and Edward (Rocky) Kolb for discussions, and the anonymous referee for constructive comments.

We acknowledge use of the Hydra cluster at IFT-UAM/CSIC (Madrid). This
research made use of Astropy,\footnote{\url{http://www.astropy.org}} a
community-developed core Python package for Astronomy
\citep{astropy:2013, astropy:2018}, TOPCAT \citep{2005ASPC..347...29T}
and STILTS \citep{2006ASPC..351..666T}. This work has made use of data from the European Space Agency (ESA)
mission \textit{Gaia} (\url{https://www.cosmos.esa.int/gaia}), processed
by the \textit{Gaia} Data Processing and Analysis Consortium (DPAC,
\url{https://www.cosmos.esa.int/web/gaia/dpac/consortium}). Funding
for the DPAC has been provided by national institutions, in particular
the institutions participating in the \textit{Gaia} Multilateral
Agreement.

FM and JGB are supported by the Research Project FPA2015-68048-C3-3-P [MINECO-FEDER] and the Centro de Excelencia Severo Ochoa Program SEV-2016-0597. DB is  been funded by the Spanish State Research Agency (AEI) Project No.ESP2017-87676-C5-1-R and No. MDM-2017-0737 Unidad de Excelencia ``Mar\'ia de Maeztu''- Centro de Astrobiolog\'ia (INTA-CSIC).



\bibliographystyle{mnras_eprint}
\bibliography{biblio}



\appendix

\section{Alternative stellar selection}
\label{sec:altercuts}

As an alternative to the stellar selection mentioned in section~\ref{sec:hvs}, here we consider 1747 stars that still satisfy cuts given in \citet{2018MNRAS.tmp.2466M}, total heliocentric velocities $v-\sigma_v \gtrsim 500$~km/s and exclude possibly contaminated radial velocities \citep{2019MNRAS.486.2618B}, but have an higher unbound probability threshold $P_{\rm ub} > 0.9$. Most of the stars have negative parallax, an indicator of poor data quality due to the absence of further photometric and astrometric cuts.

\begin{figure}
  \centering
  \includegraphics[width=\columnwidth]{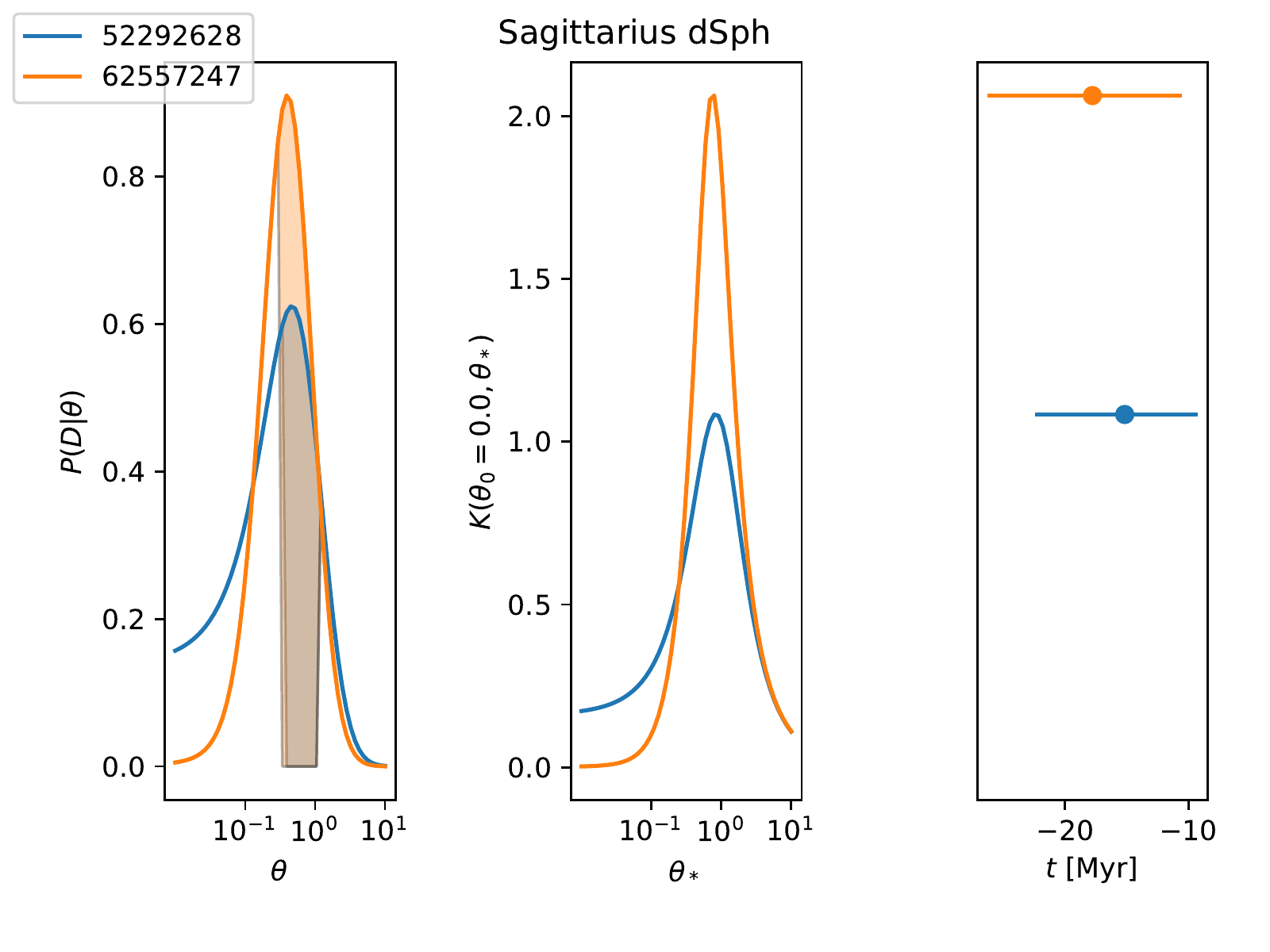}
  \caption{Same as figure~\ref{fig:sag}, but for different stars. The Bayes factor in
    the central panel is computed setting a lower prior threshold $\theta_0=0$.}
  \label{fig:sag_pub90}
\end{figure}

Assuming \citet{2018MNRAS.tmp.2466M} distances and setting a lower prior threshold $\theta_0=0$ for the computation of the Bayes factor, the same search described in section~\ref{sec:results} brings to figure~\ref{fig:sag_pub90}. Two stars are compatible with having scattered with compact objects in Sagitarius dSph 10--30~Myr ago. If we set $\theta_0=0.36$ as discussed in section~\ref{sec:results}, we find that the maximum Bayes factors increase to $K \sim 2, 10$ for stars 52292628 and 62557247, respectively, and that a third star (62779130) has positive scattering evidence ($K \gtrsim 1$). We verified that all stars are compatible with being at least $\sim 80$~Myr old. (More information about these sources is given in table~\ref{tab:hvs}.)
    
However, results are not confirmed when we repeate the same search assuming \citet{2019A&A...628A..94A} distances. Furthermore, \citet{2019A&A...628A..94A} flags the aforementioned stars as possible spurious astrometry due to large renormalised unit-weight error (RUWE). Hence, in this case positive evidence for scattering events may be driven not only by distance estimate systematic errors, but also by poor astrometric fit. 

\section{Galactocentric coordinates}
\label{sec:coord}

Galactocentric coordinates are defined as a Cartesian right-handed
system with the $x$-axis pointing from the position of the Sun
projected on the Galactic midplane to the Galactic center, the
$y$-axis roughly pointing towards Galactic longitude $\ell=90^{\circ}$
and the $z$-axis points roughly towards the North Galactic Pole (we
define the Galactic plane to be the normal to the north pole of
Galactic coordinates defined by \citet{1960MNRAS.121..164B}). The
Galactic Center right-ascension and declination are taken to be
$\alpha=17:45:37.224$ hr and $\delta=-28:56:10.23$ deg, respectively
\citep{2004ApJ...616..872R}. We assume the distance from the Sun to
the Galactic Center to be $8.33 \pm 0.35$~kpc
\citep{2009ApJ...692.1075G} and its height above the Galactic midplane
to be $27 \pm 4$~pc \citep{2001ApJ...553..184C}. Galactocentric
velocities are definied assuming a circular velocity of 220 km/s at
solar radius \citep{2015ApJS..216...29B} and a Sun peculiar velocity
with respect to the Galactic center
$(V_x, V_y, V_z)_{\sun} = (11.1 \pm 0.74 \pm 1, 12.24 \pm 0.47 \pm 2,
7.25 \pm 0.37 \pm 0.5)$~km/s with additional systematic errors
$(1, 2, 0.5)$~km/s \citep{2010MNRAS.403.1829S}.

\section{Orbits data}
\label{sec:datarefs}

Tales~\ref{tab:gc_id} and \ref{tab:gal_id} list source identifiers for
GC and galaxies used in the main analysis.  Table~\ref{tab:sag} shows
Sagittarius dSph phase-space data and size.  Table~\ref{tab:hvs} shows
phase-space data and derived quantities for \textit{Gaia} DR2 stars
compatible with having scattered with compact objects within Sagittarius dSph when assuming \citet{2018MNRAS.tmp.2466M} distances.

GC, galaxy and star catalogs are available as ancillary files at \url{https://arxiv.org/abs/1907.09298}.

\begin{table}
  \centering
  \caption{Globular clusters identifiers.}
  \label{tab:gc_id}
  \begin{tabular}{ccccc}
    \hline
NGC0104 & NGC5272 & NGC6218 & NGC6388 & NGC6656 \\
NGC0288 & NGC5286 & NGC6254 & NGC6397 & NGC6681 \\
NGC0362 & NGC5466 & NGC6266 & NGC6440 & NGC6752 \\
NGC1851 & NGC5897 & NGC6273 & NGC6453 & NGC6779 \\
NGC1904 & NGC5904 & NGC6284 & NGC6522 & NGC6809 \\
NGC2298 & NGC5986 & NGC6287 & NGC6535 & NGC6838 \\
NGC2808 & NGC6093 & NGC6293 & NGC6541 & NGC6864 \\
NGC3201 & NGC6121 & NGC6304 & NGC6544 & NGC7078 \\
NGC4372 & NGC6144 & NGC6341 & NGC6626 & NGC7089 \\
NGC4833 & NGC6171 & NGC6352 & NGC6637 & NGC7099 \\
NGC5139 & NGC6205 &   &   &   \\
\hline
  \end{tabular}
\end{table}

\begin{table*}
  \centering
  \caption{Nearby galaxy identifiers and detailed references
    (complementing those given in section~\ref{sec:gc_gal}) for
    position, size and other useful measurements. All systems are
    identified as dwarf galaxies, excluded Andromeda, Large Magellanic
    Cloud (LMC), Small Magellanic Cloud (SMC) and objects of ambiguous
    nature denoted by an asterisk (*).}
  \label{tab:gal_id}
  \begin{tabular}{p{0.15\linewidth}p{0.7\linewidth}}
    \hline
    *Draco II & \citet{2015ApJ...813...44L} \\
    *Grus 1 & \citet{2015ApJ...805..130K} \\
    *Horologium 1 & \citet{2015ApJ...805..130K, 2015ApJ...807...50B}\\
    *Reticulum 2 & \citet{2015ApJ...807...50B, 2015ApJ...805..130K} \\
    *Triangulum II & \citet{2015ApJ...802L..18L} \\
    *Tucana III & \citet{2015ApJ...813..109D} \\
    Andromeda & See section~\ref{sec:gc_gal} \\
    Antlia II & See section~\ref{sec:gc_gal} \\
    Bootes (I) & \citet{2006ApJ...653L.109D, 2007MNRAS.380..281M,
                 2008ApJ...684.1075M, 2011ApJ...736..146K, 2009ApJ...696..385G,
                 2010ApJ...723.1632N} \\
    Bootes II & \citet{2008ApJ...688..245W, 2009ApJ...690..453K,
                2008ApJ...684.1075M, 2009ApJ...696..385G} \\
    Canes Venatici (I) & \citet{2008ApJ...672L..13M, 2007ApJ...670..313S,
                         2008ApJ...684.1075M, 2009ApJ...696..385G,
                         2008ApJ...685L..43K, 2011ApJ...727...78K} \\
    Canes Venatici II & \citet{2008ApJ...675L..73G, 2007ApJ...670..313S,
                        2008ApJ...684.1075M, 2009ApJ...696..385G,
                        2008ApJ...685L..43K, 2011ApJ...727...78K} \\
    Carina & \citet{2009AJ....138..459P, 2009AJ....137.3100W,
             1995MNRAS.277.1354I, 2008ApJ...688L..75W, 2009ApJ...696..385G,
             2006AJ....131..895K} \\
    Coma Berenices & \citet{2007ApJ...654..897B, 2007ApJ...670..313S,
                     2008ApJ...684.1075M, 2009ApJ...696..385G,
                     2008ApJ...685L..43K, 2011ApJ...727...78K} \\
    Draco & \citet{2004AJ....127..861B, 2007ApJ...667L..53W,
            2008ApJ...684.1075M, 2004ApJ...611L..21W,
            2009ApJ...696..385G, 2011ApJ...727...78K} \\
    Eridanus 2 & \citet{2015ApJ...805..130K, 2015ApJ...807...50B} \\
    Fornax & \citet{2009AJ....138..459P, 2009AJ....137.3100W,
             1995MNRAS.277.1354I, 2008ApJ...688L..75W,
             2009ApJ...694L.144W, 2009ApJ...696..385G,
             2011ApJ...727...78K} \\
    Hercules & \citet{2007ApJ...668L..43C, 2009AA...506.1147A,
               2008ApJ...684.1075M, 2009ApJ...696..385G,
               2008ApJ...685L..43K, 2011ApJ...727...78K} \\
    Hydra II & \citet{2015ApJ...804L...5M, 2015ApJ...810...56K} \\
    LMC & See section~\ref{sec:gc_gal} \\
    Leo I & \citet{2004MNRAS.354..708B, 2008ApJ...675..201M,
            1995MNRAS.277.1354I, 2009ApJ...696..385G,
            2011ApJ...727...78K} \\
    Leo II & \citet{2005MNRAS.360..185B, 2007ApJ...667L..53W,
             1995MNRAS.277.1354I, 2009ApJ...696..385G,
             2011ApJ...727...78K} \\
    Leo IV & \citet{2009ApJ...699L.125M, 2007ApJ...670..313S,
             2010ApJ...710.1664D, 2009ApJ...696..385G,
             2008ApJ...685L..43K, 2011ApJ...727...78K} \\
    Leo V & \citet{} \\
    SMC & See section~\ref{sec:gc_gal} \\
    Sagittarius dSph & \citet{ 1994Natur.370..194I, 1997AJ....113..634I,
                       1998ApJ...508L..55M, 2003ApJ...599.1082M,
                       2004MNRAS.353..874M, 2007ApJ...670..346C,
                       2009ApJ...696..385G, 2011ApJ...727L...2P}\\    
    Sculptor & \citet{2008AJ....135.1993P, 2009AJ....137.3100W,
               1995MNRAS.277.1354I, 2008ApJ...688L..75W,
               1998AJ....116.1690C, 2009ApJ...696..385G,
               2009ApJ...705..328K, 2011ApJ...727...78K} \\
    Segue (I) & \citet{2007ApJ...654..897B, 2011ApJ...733...46S,
                2008ApJ...684.1075M, 2009ApJ...696..385G,
                2010ApJ...723.1632N} \\
    Segue II & \citet{2009MNRAS.397.1748B, 2013ApJ...770...16K} \\
    Sextans (I) & \citet{2009ApJ...703..692L, 2009AJ....137.3100W,
                  1995MNRAS.277.1354I, 2008ApJ...688L..75W,
                  2009ApJ...696..385G, 2011ApJ...727...78K} \\
    Tucana 2 & \citet{2015ApJ...805..130K, 2015ApJ...807...50B} \\
    Ursa Major (I) & \citet{2008AA...487..103O, 2007ApJ...670..313S,
                     2008ApJ...684.1075M, 2009ApJ...696..385G,
                     2008ApJ...685L..43K, 2011ApJ...727...78K} \\
    Ursa Major II & \citet{2006ApJ...650L..41Z, 2007ApJ...670..313S,
                    2008ApJ...684.1075M, 2009ApJ...696..385G,
                    2008ApJ...685L..43K, 2011ApJ...727...78K} \\
    Ursa Minor & \citet{2002AJ....123.3199C, 2009ApJ...704.1274W,
                 1995MNRAS.277.1354I, 2004ApJ...611L..21W,
                 2009ApJ...696..385G, 2011ApJ...727...78K} \\
    Willman 1 & \citet{2006astro.ph..3486W, 2007MNRAS.380..281M,
                2008ApJ...684.1075M, 2009ApJ...696..385G,
                2011AJ....142..128W} \\
    \hline
  \end{tabular}
\end{table*}

\begin{table}
  \centering
  \caption{Sagittarius dSph phase-space and size data. Rows
    correspond to right-ascension $\alpha$ and declination $\delta$
    (ICRS at epoch J2000.0), distance modulus $\mu$, proper motions in
    right-ascension direction
    $\mu_{\alpha*} = \mu_{\alpha}\cos{\delta}$ and in declination
    direction $\mu_{\delta}$, radial velocity along the line-of-sight
    $V_{r}$ and half-light radius measured along the major axis
    $r_h$.}
  \label{tab:sag}
  \begin{tabular}{ll}
    \hline
    $\alpha$ [deg]          & 283.8313 \\
    $\delta$ [deg]          & -30.4606 \\
    $\mu$                   & $17.13 \pm 0.11$ \\
    $\mu_{\alpha*}$ [mas/yr] & $-2.736 \pm 0.044$ \\
    $\mu_{\delta}$ [mas/yr]  & $-1.357 \pm 0.043$ \\
    $V_{r}$ [km/s]          & $140 \pm 2$ \\
    $r_h$ [arcmin]          & $342 \pm 12$ \\
    \hline
  \end{tabular}
\end{table}

\begin{landscape}
\begin{table}
  \caption{HVS compatible with having scattered with compact objects within Sagittarius
    dSph. Columns correspond to \textit{Gaia} DR2 source identifier, RA and declination, parallax, proper motion in RA direction
    $\mu_{\alpha*}=\mu_{\alpha*}\cos(\delta)$, proper motion in
    declination direction, proper motions correlation coefficient, radial velocity and heliocentric
    distance derived in \citet{2018MNRAS.tmp.2466M}, heliocentric distance derived in \citet{2019A&A...628A..94A} and probability of being unbound from the Milky Way potential. The unbound probability relies on \citet{2018MNRAS.tmp.2466M} distances. Reference epoch for astrometric quantities is J2015.5. While here we report the full source identifier, in the main text we only refer to the first 8 digits. The first two stars are discussed in section~\ref{sec:results}, the others in appendix~\ref{sec:altercuts}}
  \label{tab:hvs}
  \begin{tabular}{|r|r|r|r|r|r|r|r|r|r|r|r|r|r|}
\hline
  \multicolumn{1}{|c|}{source} &
  \multicolumn{1}{c|}{($\alpha$, $\delta$)} &
  \multicolumn{1}{c|}{$\varpi$} &
  \multicolumn{1}{c|}{$\mu_{\alpha*}$} &
  \multicolumn{1}{c|}{$\mu_{\delta}$} &
  \multicolumn{1}{c|}{$c_{\mu_{\alpha*} \mu_{\delta}}$} &
  \multicolumn{1}{c|}{$V_r$} &
  \multicolumn{1}{c|}{$d_{\rm Marchetti}$} &
  \multicolumn{1}{c|}{$d_{\rm Anders}$} &
  \multicolumn{1}{c|}{$P_{\rm ub}$} \\
  \multicolumn{1}{|c|}{} &
  \multicolumn{1}{c|}{[deg]} &
  \multicolumn{1}{c|}{[mas]} &
  \multicolumn{1}{c|}{[mas/yr]} &
  \multicolumn{1}{c|}{[mas/yr]} &
  \multicolumn{1}{c|}{} &
  \multicolumn{1}{c|}{[km/s]} &
  \multicolumn{1}{c|}{[kpc]} &
  \multicolumn{1}{c|}{[kpc]} &
  \multicolumn{1}{c|}{} \\
\hline
6198594250104473728 & (223.0436, -37.9298) & $0.10 \pm 0.04$ & $-21.14 \pm 0.06$ & $2.40 \pm 0.05$ & -0.12 & $96.0 \pm 1.3$ & $9.4_{-2.0}^{+3.3}$ & $5.8_{-1.3}^{+1.9}$ & 0.87\\
6242065813132837376 & (241.4245, -24.0908) & $0.08 \pm 0.04$ & $-11.97 \pm 0.07$ & $2.71 \pm 0.04$ & $0.03$ & $-60.5 \pm 1.3$ & $11.7_{-3.0}^{+4.8}$ & $7.2_{-1.8}^{+1.8}$ & 0.54\\
\hline
  5229262874909180928 & (162.8655, -73.4932) & $-3.79 \pm 1.14$ & $-30.45 \pm 2.16$ & $38.96 \pm 2.12$ & $-0.08$ & $51.9 \pm 1.3$ & $8.6 _{-3.2}^{+4.8}$ & $2.0_{-0.5}^{+1.8}$ & 0.97 \\
  6255724732546338688 & (228.2981, -20.7720) & $-2.59 \pm 0.73$ & $-31.37 \pm 1.42$ & $10.88 \pm 1.36$ & $-0.12$ & $-109.3 \pm 0.8$ & $8.6 _{-3.6}^{+5.3}$ & $2.3_{-0.8}^{+1.9}$ & 0.93 \\
  6277913053288720512 & (214.1732, -20.3474) & $-2.02 \pm 0.43$ & $-40.19 \pm 0.69$ & $11.60 \pm 0.54$ & $-0.22$ & $-39.7 \pm 2.9$ & $10.1_{-3.3}^{+5.5}$ & $4.3_{-1.4}^{+2.5}$ & 1.00 \\
\hline
  \end{tabular}
\end{table}
\end{landscape}


\bsp	
\label{lastpage}
\end{document}